\documentclass[pra,a4paper,twocolumn,english,showpacs,superscriptaddress]{revtex4}
\newcommand{\be}{\begin{equation}}
\newcommand{\ee}{\end{equation}}
\newcommand{\bea}{\begin{eqnarray}}
\newcommand{\eea}{\end{eqnarray}}

\usepackage{amsmath,amssymb}
\usepackage[T1]{fontenc}
\usepackage{hyperref}
\usepackage{graphicx,color}
\usepackage{bm}
\usepackage[utf8]{inputenc}
\usepackage{upgreek}
\usepackage{ulem}

\usepackage{babel}
\makeatother

\begin{document}
\title{Polarization-based Light-Atom Quantum Interface with an All-optical Trap}

\author{M. Kubasik\footnotemark[2] \footnotetext{\footnotemark[2]These authors contributed equally
to this work.}}
	\email{m.kubasik1@physics.ox.ac.uk}
	\affiliation{ICFO-Institut de Ciencies Fotoniques, Mediterranean Technology Park,  08860 Castelldefels (Barcelona), Spain}
	\affiliation{Present address: Clarendon Laboratory, Parks Road, Oxford. OX1 3PU, United Kingdom}
\author{M. Koschorreck\footnotemark[2]}
	\affiliation{ICFO-Institut de Ciencies Fotoniques, Mediterranean Technology Park,  08860 Castelldefels (Barcelona), Spain}
\author{M. Napolitano}
	\affiliation{ICFO-Institut de Ciencies Fotoniques, Mediterranean Technology Park,  08860 Castelldefels (Barcelona), Spain}
\author{S.~R.~de~Echaniz}
	\affiliation{ICFO-Institut de Ciencies Fotoniques, Mediterranean Technology Park,  08860 Castelldefels (Barcelona), Spain}
\author{H. Crepaz}
	\affiliation{ICFO-Institut de Ciencies Fotoniques, Mediterranean Technology Park,  08860 Castelldefels (Barcelona), Spain}
	\affiliation{Present address: Clarendon Laboratory, Parks Road, Oxford. OX1 3PU, United Kingdom}
\author{J. Eschner}
	\affiliation{ICFO-Institut de Ciencies Fotoniques, Mediterranean Technology Park,  08860 Castelldefels (Barcelona), Spain}
\author{E. S. Polzik}
  \affiliation{Niels Bohr Institute, Blegdamsvej 17, 2100 Copenhagen Ø, Denmark}
\author{M. W. Mitchell}
	\affiliation{ICFO-Institut de Ciencies Fotoniques, Mediterranean Technology Park,  08860 Castelldefels (Barcelona), Spain}
 
\begin{abstract}
We describe the implementation of a system for studying light-matter interactions using an ensemble of $10^6$ cold rubidium 87 atoms, trapped in a single-beam optical dipole trap. In this configuration the elongated shape of the atomic cloud increases the strength of the collective light-atom coupling. Trapping all-optically allows for long storage times in a low decoherence environment. We are able to perform several thousands of measurements on one atomic ensemble with little destruction. We report results on paramagnetic Faraday rotations from a macroscopically polarized atomic ensemble. Our results confirm that strong light-atom coupling is achievable in this system which makes it attractive for single-pass quantum information protocols.
\end{abstract}
\maketitle

\section{Introduction}\label{sec:intro}

The interaction of light and matter is of basic and practical
importance in a great variety of scientific fields.  The
interactions themselves can be studied at the most fundamental
level when the quantum character of both the light and the matter
is evident, and this has motivated much work in quantum optics. At
the same time, control of quantum light-matter interactions is a
key requirement for quantum memories \cite{julsgaard04} and
quantum networking \cite{briegel98}.  Observing the quantum
effects in both light and matter is challenging, but has been
demonstrated in a few physical systems. These include cavity
quantum electro-dynamics (QED) with individual atoms
\cite{Wilk2007,Boozer2007,Deleglise2008}, and circuit QED with individual
solid-state systems \cite{wallraff04}. Cavity-based approaches
have also been applied to ensembles containing few \cite{mielke98}
and many \cite{lambrecht96, schleier2008} atoms. Another approach
uses room temperature \cite{hald99} or laser-cooled
\cite{chaudhury06, appel08} atomic ensembles without cavities. In
these systems, a freely-propagating light beam passes through the
ensemble and the light and atoms interact during a single pass.
The quantum variables of the ensemble and light are collective
variables such as total atomic spin and Stokes operators,
respectively.  The absence of a cavity is compensated by the use
of a large number of atoms, typically $10^9$ to $10^{12}$, so it
is still possible to obtain a large net interaction and perform
quantum information tasks, e.g., quantum memory
\cite{julsgaard04}.

The use of polarized, near-resonant probes to measure the spin
state of atoms was proposed in the context of optical pumping
\cite{happer72} by Kastler \cite{kastler51} and demonstrated by
Manuel and Cohen-Tannoudji \cite{manuel63}, and has been used in a
number of contexts since then.  Modern work with cold atoms
includes probing of atoms in a MOT with a polarization-squeezed
beam \cite{sorensen98}, observation of Larmor precession due to
few-pT fields \cite{isayama99}, and estimation
and control of atomic spin states \cite{smith06,chaudhury07}.
Closely related, linear \cite{choi05} and nonlinear
\cite{budker02} magneto-optic effects have been extensively
studied for their potential application in high-sensitivity
magnetometry and measurement of fundamental symmetries.

There are several reasons to improve the technical aspects of the
light-atom interface.  Most immediately, using a non-resonant and
non-magnetic atom trap will allow longer spin coherence
times, and the strong resonant
interactions available with cold atoms.  At the same time, an
improved trap geometry is expected to increase the strength of the
collective light-atom coupling \cite{Mueller2005PRAv71p033803}.  An increase in
interaction strength offers an important practical advantage: it
should be possible to perform similar experiments, e.g., obtain
the same degree of spin squeezing \cite{de05} with far fewer
atoms. This implies a larger ratio of `quantum' spin components
(those which are required by uncertainty relations) to `classical'
components, and thus a reduced sensitivity to classical
fluctuations.  The use of nearly stationary atoms permits
interactions on time scales that are limited only by
time-bandwidth considerations of the probe light, and not by
time-scales of atomic motion \cite{Koschorreck2008}. Practically,
this means that sub-micro-second pulses can interact with the
atomic ensemble; single-photon and non-Gaussian state generation
has been demonstrated with this time-scale
\cite{neergaard-nielsen06}.

In this paper we describe an experimental system for studying
light-matter interactions using an ensemble of $\sim 10^6$ cold
rubidium-87 atoms, trapped in a single-beam optical dipole trap.
The observed trap lifetime is very long and permits thousands of 
interactions with the same sample of atoms. The trapping and
probing systems are designed to optimize the single-pass
interaction of the light. This allows us to achieve a larger
optical interaction per atom than observed in other single-pass
systems, and to make quantum non-demolition measurements of the
collective atomic spin with unprecedented precision.  Additional
features that arise from the use of cold, dipole-trapped atoms
include selection, by probe tuning, of designer Hamiltonians for
the light-matter interaction. This may enable new operations in
quantum information \cite{deechaniz08}.

The paper is organized as follows: In Part \ref{sec:physics} we review the
physical considerations in single-pass interaction of polarized
light with cold ensembles. In Part \ref{sec:atomsample} we describe the atomic
ensemble and relevant aspects of the trapping and cooling system
to produce it. The various optical pumping schemes employed in the preparation of polarized atomic ensembles are discussed in Part \ref{sec:oppump}.
Part \ref{sec:probes} concentrates on the production of polarized probe beams and their subsequent detection. In Part \ref{sec:results} we report first results on sensitive measurement of atomic polarization by Faraday rotation. Conclusions and future plans are included in Part \ref{sec:summaryoutlook}.

\section{Probing atomic spin degrees of freedom with off-resonant light}\label{sec:physics}

\begin{figure}[b!]
    \includegraphics[width=\columnwidth,keepaspectratio=true]{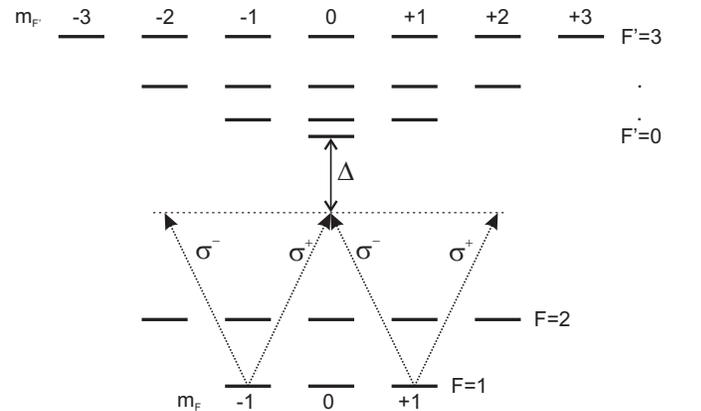}
    \caption{\label{fig:LevelScheme}Level scheme of the $^{87}$Rb D$_2$ line probed by a linearly polarized field with a detuning $\Delta$ from the $F=1 \rightarrow F'$ transitions.}
\end{figure}

We will study the interaction between an ensemble of cold $^{87}$Rb atoms prepared in the $|F=1,\;m=\pm 1\rangle$ ground states (Fig.~\ref{fig:LevelScheme}) and a linearly polarized probe pulse of duration $\tau$, tuned to the D$_2$ line, and travelling in the $z$-direction (see Fig.~\ref{fig:beam-setup}). The spin of the atoms can be described by the collective atomic pseudo-spin operators $\mathbf{\hat{J}}$
\begin{equation}\label{eq:Jdef}
\begin{split}
  \hat J_0  &= \frac{1}{2}\hat{N_a} , \\
  \hat J_{\mathbf{x}}  &= \frac{1}{2}\sum\limits_k {\left( {\hat F_{x,k}^2 - \hat F_{y,k}^2 } \right)}, \\
  \hat J_{\mathbf{y}}  &= \frac{1}{2}\sum\limits_k {\left( {\hat F_{x,k} \hat F_{y,k}  + \hat F_{y,k} \hat F_{x,k}} \right) } , \\
  \hat J_{\mathbf{z}}  &= \frac{1}{2}\sum\limits_k {\hat F_{z,k} } ,
\end{split}
\end{equation}
where $\hat{N_a}$ is the atom-number operator, $\hat F_{i,k}$ is the $i$th component of spin operator corresponding to the $k$th atom, and the sum is over all atoms. Similarly, the polarization of the probe field can be described by the Stokes operators $\mathbf{\hat{S}}$
\begin{equation}\label{eq:Sdef}
\begin{split}
  \hat S_0  &= \frac{1}{2}\left( {\hat a_ + ^{\dag} \hat a_ +   + \hat a_ - ^{\dag} \hat a_ -  } \right), \\
  \hat S_{\mathbf{x}}  &= \frac{1}{2}\left( {\hat a_ - ^{\dag} \hat a_ +   + \hat a_ + ^{\dag} \hat a_ -  } \right), \\
  \hat S_{\mathbf{y}}  &= \frac{i}{2}\left( {\hat a_ - ^{\dag} \hat a_ +   - \hat a_ + ^{\dag} \hat a_ -  } \right), \\
  \hat S_{\mathbf{z}}  &= \frac{1}{2}\left( {\hat a_ + ^{\dag} \hat a_ +   - \hat a_ - ^{\dag} \hat a_ -  } \right),
\end{split}
\end{equation}
where $\hat a_\pm^{\dag}$ $(\hat a_\pm)$ are the creation (annihilation) operators of the $\sigma^\pm$ modes of the field. The upright subscripts $\left\lbrace \mathbf{x}, \mathbf{y}, \mathbf{z} \right\rbrace $ distinguish non-spatial coordinates for pseudo-spin and Stokes operators from space-like coordinates, e.g., angular momentum operators. 

In general the interaction Hamiltonian consists of three terms, respectively proportional to the scalar, vectorial and tensorial part of the atomic polarizability \cite{de05, Geremia2006PRAv73p042112}. For the range of detunings used in our measurements, the tensorial term of the polarizability is at least an order of magnitude smaller than the vectorial one and can be neglected \cite{deechaniz08}. The scalar term is state-independent and therefore can be dropped entirely. The Hamiltonian then reduces to
\begin{equation}\label{eq:Hamiltonian}
    \hat{H}_I  =\hbar\frac{G}{\tau} \hat{S}_{\mathrm{z}} \hat{J}_{\mathrm{z}},
\end{equation}
where the coupling constant $G$ contains the vectorial part of the atomic polarizability $\alpha^{(1)}$ and the interaction area $A$ \cite{deechaniz08}
\begin{equation}
    G(\Delta,A) = \frac{1}{A}\frac{{\Gamma \lambda^2}}{{16 \pi}} \left(-4\delta_{0}\!\left(\Delta\right)-5\delta_{1}\!\left(\Delta\right)+5\delta_{2}\!\left(\Delta\right)\right).
\end{equation}
$\lambda$ is the  transition wavelength, $\Gamma$ is the excited state decay rate, and $\Delta$ is the probe detuning. The functions $\delta_{F'}(\Delta)=\left(\Delta+\Delta_{0,F'}\right)^{-1}$ include the finite hyperfine splittings in the excited state: $\Delta_{0,F'}$ is the hyperfine level spacing between $F'=0$ and $F'=1,2$.

To first order, and after an interaction time $\tau$, this Hamiltonian will produce the following input/output relations
\begin{equation}\label{eq:InOut}
\begin{split}
    \hat S_{\mathbf{y}}^{\text{(out)}}  &\approx \hat S_{\mathbf{y}}^{\text{(in)}}  +
        G \hat J_{\mathbf{z}}^{\text{(in)}} \hat S_{\mathbf{x}}^{\text{(in)}} , \\
    \hat S_{\mathbf{z}}^{\text{(out)}}  &= \hat S_{\mathbf{z}}^{\text{(in)}} , \\
    \hat J_{\mathbf{y}}^{\text{(out)}}  &\approx \hat J_{\mathbf{y}}^{\text{(in)}}  +
        G \hat S_{\mathbf{z}}^{\text{(in)}} \hat J_{\mathbf{x}}^{\text{(in)}} , \\
    \hat J_{\mathbf{z}}^{\text{(out)}}  &= \hat J_{\mathbf{z}}^{\text{(in)}} ,
\end{split}
\end{equation}
where $\hat{S}_{\mathbf{x}}$ is rotated onto $\hat{S}_{\mathbf{y}}$ by an amount that is proportional to $\hat{J}_{\mathbf{z}}$. This entangles the atomic and the light variables \cite{Kupriyanov2005PRAv71p032348}. With both $\mathbf{\hat{J}}$ and $\mathbf{\hat{S}}$ initially prepared in coherent states pointing in the $\mathbf{x}$-direction the average values of $\hat{J}_{\mathbf{z}}$ and $\hat{S}_{\mathbf{z}}$ are zero and the subsequent measurement of $\hat S_{\mathbf{y}}$ will result in a reduction of the variance, var($\hat{J}_{\mathbf{z}}$), below the standard quantum limit \cite{Kuzmich1998EPLv42p481, Kuzmich2003IBQuantuminformationwithp231, Thomsen2002PRAv65p061801} thus producing a squeezed state of the pseudo-spin. It follows from Eqs. \eqref{eq:Jdef}, \eqref{eq:Sdef} and \eqref{eq:InOut} that
\begin{equation}\label{eq:VarInOut}
\begin{split}
\mathrm{var}(\hat S_{\mathbf{y}}^{\text{(out)}}) &= \mathrm{var} (\hat S_{\mathbf{y}}^{\text{(in)}}) + G^2 \langle \hat S_{\mathbf{x}}^{\text{(in)}}\rangle^2 \mathrm{var} (\hat J_{\mathbf{z}}^{\text{(in)}}) \\ 
	&= \frac{N_p}{4} + G^2 \frac{N_p^2}{4} \frac{N_a}{4},
\end{split}
\end{equation}
where $N_a$ and $N_p$ are the atom and photon numbers, respectively. 

The attainable degree of spin squeezing \cite{Duan2000PRLv85p5643, Hammerer2004PRAv70p044304} depends on the coupling constant $G$, the number of atoms and the number of photons.
In the case of a small atomic sample in a dipole trap the overlap of the probe beam and the atomic sample affects each of these parameters. Therefore in this configuration the probe-sample matching is of fundamental importance \cite{Mueller2005PRAv71p033803}. Apart from the inhomogeneous coupling, in the experiment, unwanted noise is always present and this can be either due to atoms or light. These and related issues have been addressed theoretically in Refs. \cite{Duan2002PRAv66p023818, Madsen2004PRAv70p052324, de05, Koschorreck2008}. A general criterion that can be used to verify spin squeezing for a two pulse experiment in the presence of noise will be presented elsewhere \cite{QND_char_paper}.

In an experimental context it is desirable to devise a method to directly measure $G$. This can be done if a different initial state is used. For the atomic pseudo-spin polarized along the $\mathbf{z}$-axis and light polarized along the $\mathbf{x}$-axis, Eqs.~\eqref{eq:InOut} imply a rotation of the Stokes vector in the Poincar{\'e} sphere by an angle
\begin{equation}\label{eq:RotAng}
\theta = G \langle\hat J_{\mathbf{z}}\rangle = \frac{G N_a}{2}.
\end{equation}

Hence, a measurement of the Faraday rotation angle, $\theta$, provides information about $G$.
As noted earlier the effective values of $G$ and $N_a$ depend on the size and the overlap of the probe and the sample. 

Another measure of the strength of the atom-light interaction is the on-resonance optical depth
\begin{equation}\label{eq:ODdef}
 \mathrm{OD} = \sigma_0 \frac{N_a}{A},
\end{equation}
where $\sigma_0$ is the on-resonance scattering cross section.
Combining Eqs. (\ref{eq:RotAng}) and (\ref{eq:ODdef}) we obtain an expression that can be used to determine $\mathrm{OD}$ from the same Faraday rotation measurement
\begin{equation}\label{eq:ODtheta}
\mathrm{OD} = \frac{2 \sigma_0}{\widetilde{G}} \theta,
\end{equation}
where $\widetilde{G} = A G$ depends only on atomic quantities and can be readily calculated.

Atom-light entanglement and production of spin squeezed states as described by Eq.~\eqref{eq:VarInOut} are examples of the possible applications of the QND Hamiltonian \eqref{eq:Hamiltonian}. In a general situation
where the detuning $\Delta$ is allowed to take an arbitrary value, Eq. \eqref{eq:Hamiltonian} is no longer valid and the general form of the Hamiltonian has to be used \cite{de05}. This has numerous applications, including atom-number measurement, quantum cloning and quantum memory \cite{deechaniz08}. Among these the first one is especially relevant in the context of this paper. The Hamiltonian that allows for atom-number measurements is obtained for a specific value of $\Delta=462\,\mathrm{MHz}$ and is given by
\begin{equation}\label{eq:number-meas}
    \hat H_{I} \propto \alpha^{(2)}
        \left( {\hat S_{\mathbf{x}} \hat J_{\mathbf{x}} + \hat S_{\mathbf{y}} \hat J_{\mathbf{y}}} \right).
\end{equation}
For a circularly polarized probe and the pseudo-spin polarized along the $\mathbf{x}$-axis, it describes a rotation of $\hat{S}_{\mathbf{z}}$ onto $\hat{S}_{\mathbf{y}}$ proportional to $\hat{J}_{\mathbf{x}}$. 
This should prove useful because the result can be compared against the number of atoms measured by more conventional methods as for instance absorption imaging or fluorescence measurement after recapturing atoms in the MOT.

\section{Atom Trap}\label{sec:atomsample}
An ensemble of $^{87}$Rb is prepared in a double magneto-optical trap (MOT). The first stage of the MOT confines atoms only in 2 dimensions letting them move to the second stage located below, where they are trapped in 3 dimensions. This transfer is aided with a weak auxiliary beam which results in a rate of about $10^7$ atoms per second. A detailed description of the MOT apparatus can be found in Refs. \cite{Schulz2002PHD} and \cite{Crepaz2006PHD}. From the 3D MOT the cold sample is transferred to a far off-resonance dipole trap (FORT). The 2-stage configuration of the MOT with pressure in the bottom stage lower by two orders of magnitude provides fast loading and a long lifetime of the dipole trap.

Transfer to the dipole trap does not involve moving the atoms in space which reduces possible heating and losses. Instead, the FORT and the MOT are overlapped. The transition between these two traps is supported by a molasses phase: during the last 25~ms of the MOT loading the axial gradient of the magnetic field is decreased from 30 to 20~G/cm, the detuning of the cooling light is ramped from $-1.3$ $\Gamma$ to $-4$ $\Gamma$, and the repumping beam is attenuated; later the magnetic field is turned off, the repumper is blocked and the cooling light is kept on for 85~ms detuned by $-14$ $\Gamma$. This last phase leaves the atoms in the $F=1$ hyperfine level which reduces the losses due to spin-changing collisions and is also the required initial state for the following optical pumping (see \ref{sec:oppump}). In this way we fill the dipole trap with about $1.2\times 10^6$ atoms and the entire cycle takes only about 3 seconds.

The FORT is realized with a red-detuned ($1030\,\mathrm{nm}$), linearly polarized  beam, focused to a $50\,\upmu\mathrm{m}$ waist by an achromatic lens that is also used to focus the probe beams on the sample (L1 in Fig. \ref{fig:beam-setup}).
A thin disk, Yb:YAG laser provides 7~$\mathrm{W}$ of continuous power in the trap. Trapping light is brought to the apparatus by a single mode photonic-crystal fiber to assure a pure Gaussian beam and to reduce pointing instabilities. The AC~Stark shift induced on the ground states by the dipole light corresponds to a confining potential of about $260\,\upmu\mathrm{K}$ depth.

The number of atoms in the FORT is measured by switching it off and recapturing the atoms in the 3D MOT. The fluorescence signal from the MOT is recorded during $100\,\mathrm{ms}$. With the 2D MOT switched off this time is short enough to avoid capturing any atoms that are not from the dipole trap. Fluorescence light is collected with an $\mathrm{NA}=0.33$ objective \cite{Crepaz2006PHD} and sent onto a calibrated, amplified photodiode which completes the measurement. Recapture after varying dipole trapping periods allows us to investigate trap losses. 
We observe a mainly super-exponential, density-dependent decay, presumably due to two-body collisions in the trap \cite{Kuppens2000PRAv62p013406,Grimm2000AAMOPv42p95}. Fitting data after up to 90 seconds of trapping we can claim that the collision rate parameter (volume-independent) $\beta$ is close to $8\times 10^{-14}\,\mathrm{cm}^{3}\mathrm{s}^{-1}$, in agreement with other predicted and measured values for $^{87}$Rb under similar conditions \cite{Nesnidal2000PRAv62p030701(R),Beijerinck2000}. In comparison, losses due to collisions with hot background atoms are negligible. In this situation inferring the lifetime is difficult. Nonetheless, the fits suggest a value greater than 1000 seconds, which is consistent with a background pressure of $10^{-11}\,\mathrm{mbar}$ \cite{Bali1999}.

For imaging purposes the dipole-trapped sample is illuminated with light from the MOT, red-detuned by 1~$\Gamma$. A CCD camera records a fluorescence image during $100\,\upmu\mathrm{s}$. This allows us to estimate the size of the sample and the temperature of the atoms by using the time of flight (TOF) technique \cite{PhysRevLett.61.169, Adams1997KEYProg.Quant.Electr.v21p1}. To ensure that only the dipole-trapped atoms contribute to the images the trapping time in this case is set to $300\,\mathrm{ms}$. After this time, the FORT is switched off and an image of the cloud is acquired after a variable time of free expansion (up to $4\,\mathrm{ms}$). The fitted size of the cloud is $8.5\,\mathrm{mm}$ by approximately  $20\,\upmu\mathrm{m}$ (FWHM) indicating an atomic density at the center of approximately $5\times 10^{11}\,\mathrm{atoms/cm}^3$. From the time dependence of the cloud radius we infer the temperature, $25.0\pm 0.5$~$\upmu$K.

\section{State Preparation by Optical Pumping}\label{sec:oppump}
It was explained in Part II how a QND-type measurement can be used to generate spin squeezing. This scheme works with atoms initially prepared in a coherent spin state, with the pseudo-spin polarized along the $\mathbf{x}$-axis. This state is equivalent to a coherent superposition of the two extreme ground states $|F=1,\;m= 1\rangle$ and $|F=1,\;m=- 1\rangle$, (Fig. \ref{fig:LevelScheme}). On the other hand, Faraday rotation measurements that provide a method to assess the interaction strength available in the experiment are ideally performed with only one of these two states populated.
In this section we present the configuration of the pumping beams used in both situations.
\begin{figure}[htp]
	\centering
	\includegraphics[width=.45\textwidth]{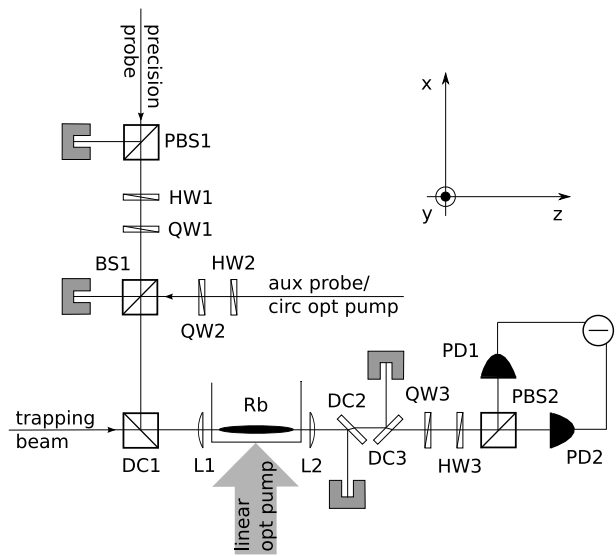}
	\caption{Simplified schematic of the setup. PBS1, PBS2 polarizing beam splitters. BS1 a 50/50 beam splitter. DC1 dichroic cube. DC2, DC3 dichroic mirrors. HW half waveplates, QW quarter waveplates. L1, L2 respectively focusing and re-collimating lenses. PD1, PD2 photodiodes.}
	\label{fig:beam-setup}
\end{figure}

Since the vacuum apparatus has been designed so that optical pumping fields can be applied from three mutually orthogonal directions \cite{Schulz2002PHD, Crepaz2006PHD}, a variety of quantum states can be prepared. With the $z$-axis acting as a quantization axis (see Fig. \ref{fig:beam-setup}), pumping atoms into either $|F=1,\;m= 1\rangle$ or $|F=1,\;m=- 1\rangle$ requires circularly polarized light propagating along $z$. The desired $m$-state is then obtained by selecting the correct helicity. We combine the pumping beam with a precision probe beam (cf. Part \ref{sec:probes}) on a 50/50 beam splitter (BS1) and set the angles of waveplates HW2 and QW2 so that  correct circular polarization in the trap results. In practice polarization is measured between the vacuum cell and the collimating lens L2. A set of two waveplates is necessary because the $s$- and $p$-polarized components acquire a different phase shift upon reflection from the dichroic cube, DC1. For the measurements presented in Part \ref{sec:results} we pump atoms with approximately 3~$\upmu$W for 50~$\upmu$s with light resonant from $F=1$ to $F'=2$ and a beam about four times as big as the transverse dimension of the atomic ensemble. At the same time we
apply light fields through the MOT beams resonant to the transition  $F=2$ to $F'=2$. This prevents atoms from being accumulated in the $F=2$ manifold. All the resonance frequencies referred to are the free space values.

 We are not compensating for light shifts from the dipole trap laser (around $\mathrm{12\,MHz}$) at this stage. In addition, a small guiding magnetic field of 0.5 G is applied in order to prevent precession of the state about any remaining stray magnetic field.

The production of coherent superposition states requires pumping light that is linearly polarized in the $xy$-plane. When such polarized light is used the atoms end up in one of the two $|F=1,\;m= 1\rangle \pm |F=1,\;m=- 1\rangle$ superpositions with the sign depending on whether the actual polarization vector points along the $x$ or $y$-axis. In our implementation the pumping beam travels along the vertical axis, $x$, and is beforehand expanded in the $z$-direction so that it better matches the elongated atomic sample. This particular configuration makes use of the lower optical depth in the propagation direction of the optical pumping beam.

\section{Probing and Detection System}\label{sec:probes}
Two probe beams have been implemented, a linearly polarized precision probe that is shot noise limited in polarization and an auxiliary probe whose polarization can be set arbitrarily. The precision probe yields a polarization rotation signal according to Eq.~\eqref{eq:RotAng}. The probe light is produced by a commercial extended cavity diode laser. Its frequency is locked to the frequency of the repumper laser of the MOT using a computer controlled electronic offset-lock on the basis of a digital PLL (ADF4252) \cite{guenter03}. This configuration allows for the detuning $\Delta$ (Fig. \ref{fig:LevelScheme}) to be varied continuously from -0.2~GHz up to -2.8~GHz. With the help of an acousto-optic modulator (AOM) rectangular pulses as short as 100~ns can be created. The beam is brought to the trap by a single-mode polarization maintaining fiber and a coherent polarization state is prepared with a thin-film polarizer of extinction ratio of $10^{5}:1$.

The main application of the auxiliary probe is an atom number measurement based on the Hamiltonian \eqref{eq:number-meas}. To this end its polarization is made circular and the atoms are prepared in the same superposition state as required for spin-squeezing experiments (see Part \ref{sec:oppump}).
The fact that the ensemble can be prepared in the same way in both cases implies that the two measurements, the QND measurement and the measurement of the number of atoms, can be performed nearly simultaneously on the same sample by sending pulses of the two probes closely separated in time. As in the case of the precision probe, pulses are produced with an AOM and the beam is fiber coupled to the trapping setup. The two waveplates HW1 and QW1 are set such as to achieve the required polarization state of the precision probe in the trap. The same is achieved for the auxiliary probe by HW2 and QW2. 
Since the auxiliary probe and the circular optical pumping (cf. Part \ref{sec:oppump}) are not used simultaneously they share the same optical path and are combined with the precision probe and the trapping beam on BS1 and DC1, respectively. In this way all the beams (with the exception of the linearly polarized optical pumping beam) propagate in a collinear fashion. Among other advantages, the collinear geometry makes it possible to use a single achromatic lens (L1) to produce the trapping potential and to focus the probe and pump beams on the atomic sample. In our case the focal length is $f = 80$~mm. The required size of the foci is achieved by adjusting the size of each of the collimated beams separately before they reach the focusing lens. We set the waist (where intensity drops by a factor of $e^{2}$) of both Gaussian probes to $w_{0} = 20 \ \upmu$m which is close to the optimum predicted in Ref. \cite{Mueller2005PRAv71p033803}. 

After passing through the vacuum cell, all the beams are again collimated with another $f = 80$~mm achromat (L2) arranged in a unit-magnification telescope configuration with the focusing lens (L1). The trapping beam is then filtered out by a pair of dichroic mirrors and the transmitted probe beams are directed onto the detection system. Before the detection, there is a quarter waveplate whose angle is set such as to compensate for the birefringence of the two dichroic mirrors at the probe wavelength. Detection is accomplished with a half waveplate that rotates the plane of polarization by $45^{\circ}$ and a polarizing beam splitter that separates the original $45^{\circ}$ and $-45^{\circ}$ components thus completing the measurement of $S_{\mathbf{y}}$.

The intensities of the two resulting beams are subtracted in a home-built differential photodetector. It consists of two main stages. The first is a charge-sensitive amplifier that integrates the difference of the two photocurrents over the duration of the optical pulse. The second is a Gaussian filter which differentiates and amplifies the integrated signal. At the end the signal is recorded by a digital storage oscilloscope and processed later on. The measured electronic noise is equivalent to the shot noise of a pulse consisting of about $10^{5}$ photons. A full description of the detector will be given elsewhere \cite{Windpassinger2008MSTvp}.

\section{Paramagnetic Faraday Rotation Measurements}\label{sec:results}
This section presents experimental results on paramagnetic Faraday rotations which are
used to determine the amount of interaction between probe beam and atomic ensemble. As described in Sec. \ref{sec:oppump}, the initial state for this measurement is either  $|F=1,\;m= 1\rangle$ or $|F=1,\;m=- 1\rangle$, i.e., a macroscopic polarization along the $\mathbf{z}$ component of the pseudo-spin.  The guiding magnetic field along $z$ is applied during pumping and probing. The rotation signal is measured by probing the sample with 1~$\upmu$s long pulses of about $4\times10^6$ photons per pulse and a period of 20~$\upmu$s.

\begin{figure}[t!]
\centering
\includegraphics[width=\columnwidth,keepaspectratio=true]{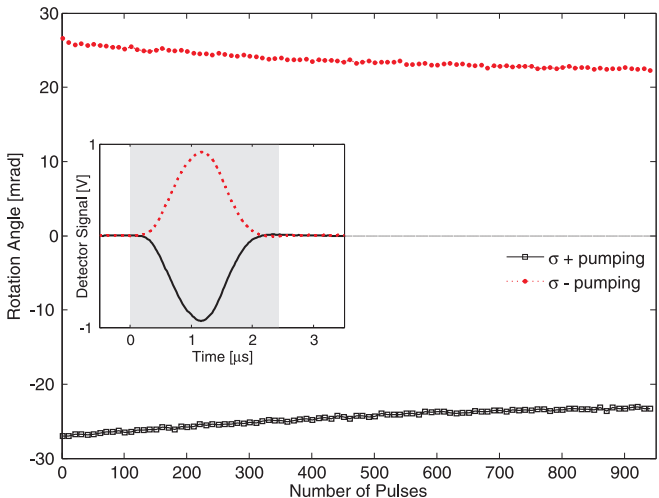}
\caption{(Color online) Rotation signal for an atomic state polarized parallel and
anti-parallel to $z$. The number of photons per pulse is $4.3\times10^6$ and the detuning
is $-1.6\,\mathrm{GHz}$. Each point in the graph represents the average value over 20 experimental runs. For clarity only every tenth point has been plotted.
Inset: Individual pulses from balanced detector. The gray area marks the integration
window.}
\label{fig:pmRotation}
\end{figure}

In a first set of  measurements we prepare atoms in either of the above-mentioned states.  The resulting rotation
signals are shown in Fig. \ref{fig:pmRotation}. The observed signals show opposite signs,
i.e., the linear polarization is tilted clockwise or anti-clockwise, respectively,  as
the light travels through the medium. The amount of rotation is the same for both states
within 2$\%$. It demonstrates that the setup is capable of producing and detecting macroscopically 
polarized atomic states. However, the degree of optical pumping, i.e., the purity of the
atomic state would have to be measured by other techniques, e.g., spin state tomography. 

The measurement of the polarization state of the atoms is
highly sensitive while producing little change in the atomic state. Each point in Fig. \ref{fig:pmRotation} 
represents a pulse of about $4 \times 10^6$ photons interacting with the atomic sample.  In
total 1000 pulses are sent, producing a decrease of signal of $< 10\%$.  Also, as seen in
that figure, the signal-to-noise ratio (SNR) is large, about 200.   Together, these
indicate that the system provides sufficient interaction for sensitive, non-destructive
measurements.  A full analysis of the information/disturbance trade-off will be the subject
of a future work \cite{Koschorreck}.

The inset of Fig. \ref{fig:pmRotation} shows the recorded Gaussian pulse shape as it is read
from the balanced detector. Our signal, which is the photon-number imbalance, $\Delta N_L$,  is calculated as the integral over the gray shaded area. The conversion factor was determined beforehand by sending a known amount of photons onto only one photodiode. For each pulse we monitor the number
of photons, $N_L$,  entering the atomic cloud.  By knowing the transmission of DC2 and DC3 we can calculate the angle of polarization rotation $\theta$ by
\begin{equation}\label{eq:thetafromExp}
   \theta = \frac{\Delta N_L'}{N_L t_h t_v},
\end{equation}
where $\Delta N_L'$ is the photon number difference actually measured and  $t_v$ and $t_h$ the amplitude transmission probability for vertical and horizontal polarization, respectively. 

\begin{figure}[t!]
\centering
\includegraphics[width=\columnwidth,keepaspectratio=true]{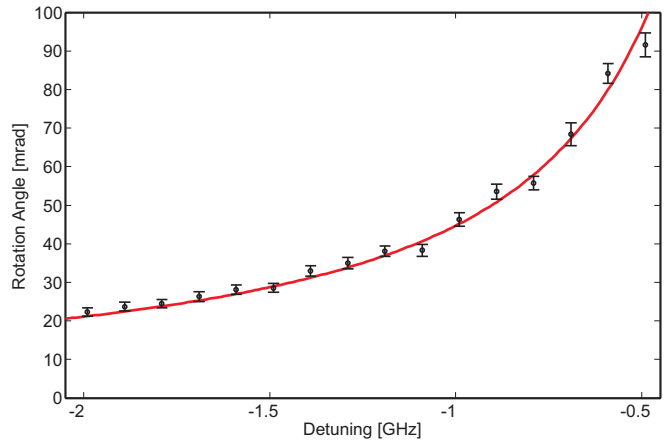}
\caption{(Color online) Paramagnetic Faraday rotation signal measured on an atomic
ensemble of about $1\times 10^6$ atoms. The detuning is measured from the resonance $F=1
\rightarrow F'=0$. The number of photons per pulse is $4 \times 10^6 $. For more details,
see text.}
\label{fig:Rotangle}
\end{figure}

In the second set of measurements we pump atoms into the state $|F=1,\;m=-1\rangle $ and
vary the detuning of the probe laser over $1.5\,\mathrm{GHz}$. This enables us to measure the dependence on detuning of the rotation angle in Eq. (\ref{eq:RotAng}). Furthermore,
we can deduce the column density of the light-atom interface and the on-resonant optical
depth. The rotation angles are plotted in Fig. \ref{fig:Rotangle}, where a single point
corresponds to an average over 40 realizations of a dipole trapped sample. Each sample
is probed by 10 pulses to determine the rotation angle. The error bars correspond to one
standard deviation.
The solid line in Fig. \ref{fig:Rotangle} is a least-square fit using Eq. \eqref{eq:RotAng}. The only free parameter is the column density $ n_c=\frac{N_a}{A}$ and
we can write the fit function as
\begin{equation}\label{eq:fitfunct}
   \theta(\Delta) = n_c \frac{\widetilde{G}(\Delta)}{2}.
\end{equation}
From the fitting we find $n_c=2.65(7)\times10^{14}$~$\mathrm{m}^{-2}$. We interpret the
measured column density as an effective number. This is to say, we assume $N_a$ atoms
homogeneously distributed over an effective area $A_{\mathrm{eff}}$ and a light beam of
the same size.  We use the effective column density and determine an effective on-resonance optical depth of $51 \pm 1$.
The term "on-resonance" here requires some explanation.  Unlike the ideal spin-1/2,
two-level atom, our atom has three resonances which each make a contribution, both to the
absorption and the optical rotation effects.  To define an "on-resonant" scattering
cross-section, we sum the scattering cross sections for the three transitions at their
respective line centers and obtain $\sigma_0 = \lambda^2/\pi$.  This cross section
accurately describes the transition when the upper hyperfine splitting can be neglected,
for example far from resonance.

The obtained value of the optical depth is very encouraging for future experiments towards spin-squeezing. Nevertheless, we are aware of the fact that our multilevel atomic system is very different from the ideal spin 1/2 atom in Ref. \cite{Kuzmich1998EPLv42p481}. Therefore, any predictions about the degree of spin squeezing achievable in our system as in Ref. \cite{Hammerer2004PRAv70p044304} would require a more complex analysis \cite{Koschorreck}. As a first hint, however, we can state the number of photons needed to observe atomic projection noise over light shot noise. If we use Eq. \ref{eq:VarInOut} and say we want to amplify the atomic over the light noise by a factor of $a$ we need a number of photons per pulse which is given by $N_L=a\frac{N_a}{\theta^2}$. $N_L$ has a quadratic dependence on the detuning which is compensated by the fact that the destruction of the atomic state scales inversely proportional to the square of the detuning. If we take the data from Fig. \ref{fig:pmRotation} and want to achieve an $a=1$, we have to use $10^9$ photons which corresponds to around 300 pulses under the used conditions.

\section{Conclusion}\label{sec:summaryoutlook}
We have presented an experimental setup for applications in continuous variables quantum information. The system consists of an ensemble of cold atoms in a red-detuned dipole trap interacting with an off-resonant probe. In order to characterize the strength of the atom-light interaction we have performed polarization rotation measurements varying the detuning of the probe over 1.5~GHz. The results obtained are in very good agreement with theoretical predictions. The very small observed destruction of the atomic state combined with the strong interaction, that in these measurements corresponds to an optical depth of 51$\pm$1, confirm that this system meets the requirements to successfully demonstrate spin squeezing. In a broader context our measurements indicate that cold atoms in far off-resonance dipole traps can provide strong interaction without the use of a cavity and that they may constitute a very promising physical system for quantum information protocols.

\begin{acknowledgements}
    The authors would like to thank P.J. Windpassinger and J.H. M\"{u}ller for their support on the differential photodetector. This work was funded by the Spanish Ministry of Science and Education under the LACSMY project (Ref. FIS2004-05830) and the Consolider-Ingenio 2010 Project ``QOIT''.
\end{acknowledgements}

\bibliographystyle{prsty}
\bibliography{NGAEbib}

\end{document}